\documentclass[12pt,a4paper]{article}
\usepackage{cite}
\usepackage{amsmath,amssymb,graphicx}
\usepackage{float}

\numberwithin{equation}{section}
\allowdisplaybreaks[3]

\newcommand{\al}{\alpha}
\newcommand{\ep}{\epsilon}

\newcommand{\La}{\Lambda}

\newcommand{\lb}{\lbrack}
\newcommand{\rb}{\rbrack}

\newcommand{\msc}[1]{\mbox{\scriptsize #1}}
\newcommand{\dsp}{\displaystyle}
\newcommand{\scs}[1]{{\scriptstyle #1}}

\newcommand{\bm}[1]{{\boldmath #1}}

\newcommand{\bc}{\Bbb C}

\newcommand{\bz}{\Bbb Z}

\newcommand{\bh}{\Bbb H}

\newcommand{\bsz}{\Bbb Z}

\newcommand{\cJ}{{\cal J}}

\newcommand{\cN}{{\cal N}}
\newcommand{\cM}{{\cal M}}
\newcommand{\cF}{{\cal F}}

\newcommand{\cQ}{{\cal Q}}

\newcommand{\tL}{\tilde{L}}
\newcommand{\tJ}{\tilde{J}}

\newcommand{\hc}{\hat{c}}

\newcommand{\hgamma}{\widehat{\gamma}}
\newcommand{\hdelta}{\widehat{\delta}}

\newcommand{\Th}[2]{\Theta_{#1,#2}}
\renewcommand{\th}{{\theta}}

\newcommand{\ch}[2]{\mbox{ch}^{#1}_{#2}}

\newcommand{\tr}{\mbox{Tr}}
\renewcommand{\mod}{\mbox{mod}}

\newcommand{\nn}{\nonumber\\}

\newcommand{\NS}{\mbox{NS}}
\newcommand{\tNS}{\widetilde{\mbox{NS}}}
\newcommand{\R}{\mbox{R}}
\newcommand{\tR}{\widetilde{\mbox{R}}}
\newcommand{\sNS}{\msc{NS}}
\newcommand{\stNS}{\widetilde{\msc{NS}}}
\newcommand{\sR}{\msc{R}}
\newcommand{\stR}{\widetilde{\msc{R}}}

\newcommand{\any}{{}^{\forall}}
\newcommand{\ex}{{}^{\exists}}

\newcommand{\bigbox}[2]{{\scs{#1}} \hspace{0.3mm}  \raise-1mm\hbox{$ \underset{#2} {\scalebox{1.9}{\mbox{$\Box$}}} $}}

\newcommand{\eqn}[1]{(\ref{#1})}

\begin{document}


\begin{titlepage}
 \
 \renewcommand{\thefootnote}{\fnsymbol{footnote}}
 \font\csc=cmcsc10 scaled\magstep1
 {\baselineskip=16pt
  \hfill
}

 \baselineskip=20pt
\vskip 1cm
 
\begin{center}

{\bf \Large 

Non-SUSY Heterotic String Vacua of Gepner Models with Vanishing Cosmological Constant

} 

 \vskip 1.2cm

\noindent{ \large Koji Aoyama}\footnote{\sf ro0018hk@ed.ritsumei.ac.jp},

\vspace{2mm}


\noindent{ \large Yuji Sugawara}\footnote{\sf ysugawa@se.ritsumei.ac.jp},

\medskip

 {\it Department of Physical Sciences, 
 College of Science and Engineering, \\ 
Ritsumeikan University,  
Shiga 525-8577, Japan}

\end{center}

\bigskip

\begin{abstract}

We study a natural generalization of that given in \cite{AS} to heterotic string. 
Namely, starting from the generic Gepner models for Calabi-Yau 3-folds, we construct the
non-SUSY heterotic string vacua with the vanishing cosmological constant at the one loop. 
We especially focus on the asymmetric orbifolding based on some discrete subgroup of the chiral $U(1)$-action
which acts on both of the Gepner model and the $SO(32)$ or $E_8\times E_8$-sector. 
We present a classification of the relevant orbifold models leading to   
the string vacua with the properties mentioned above. 
In some cases, the desired vacua can be constructed in the manner quite similar to those given in 
\cite{AS} for the type II string, in which the orbifold groups contain two generators with the discrete torsions. 
On the other hand, we also have simpler models that are just realized
as the asymmetric orbifolds of cyclic groups with only one generator.

\end{abstract}

\setcounter{footnote}{0}
\renewcommand{\thefootnote}{\arabic{footnote}}

\end{titlepage}

\baselineskip 18pt

\vskip2cm 
\newpage


\section{Introduction and Summary}

It has been an interesting subject in superstring theory 
to explore the non-supersymmetric vacua with 
the vanishing cosmological constant 
(at the level of one-loop, at least), probably motivated by the cosmological constant problem.  
The consistent type II string vacua with such non-trivial property has been first constructed in 
\cite{Kachru1,Kachru2,Kachru3} 
based on some non-abelian orbifolds of higher dimensional tori, followed by studies {\em e.g.} in 
\cite{Harvey,Shiu-Tye,Blumenhagen:1998uf,Angelantonj:1999gm,Antoniadis,Aoki:2003sy}. 
More recently, several non-SUSY vacua with this property have been constructed
as the asymmetric orbifolds \cite{Narain:1986qm} by simpler cyclic groups 
in \cite{SSW,SWada}.


On the other hand, in heterotic string theory, there have been the studies of the string vacua with 
cosmological constants exponentially suppressed with respect to some continuous parameter expressing the `distance' 
from the SUSY vacua 
(for instance, the radii of tori of compactifications as given in \cite{SS}) 
in  \cite{IT,Harvey}, and more recently, {\em e.g} in \cite{Blaszczyk:2014qoa,Angelantonj:2014dia,Faraggi:2014eoa,Abel:2015oxa,
Kounnas:2015yrc,Abel:2017rch,GrootNibbelink:2017luf,Basile:2018irz,ItoyamaN}  from the various viewpoints of model building in string phenomenology. 


Now, the purpose of current study is to construct the non-SUSY heterotic string vacua with the vanishing cosmological constant
at one-loop based on the {\em non-toroidal\/} models. 
The method we adopt is a natural generalization of those given in our previous work \cite{AS}. 
That is, we start from the generic Gepner models \cite{Gepner} for Calabi-Yau 3-folds, and 
construct the non-SUSY heterotic string vacua by implementing some asymmetric orbifolding. 
Since we have various $U(1)$-symmetries in the Gepner model as well as the $SO(32)$ or $E_8\times E_8$-sector
in the left-mover (which we assume bosonic), it would be quite natural to make the orbifolding associated to some cyclic  
subgroup of these $U(1)$-actions. 
Indeed, let us denote the generator of such a cyclic subgroup as `$\delta_L$'. Then, 
it is possible to construct the non-SUSY string vacua by making the asymmetric orbifolding defined by the operator 
$$
\hdelta \equiv \delta_L \otimes (-1)^{F_R},
$$ 
where $F_R$ denotes the space-time fermion number (in other words, $(-1)^{F_R}$ acts as the sign-flip on the right-moving Ramond sector). 
It is obvious that the orbifold projection generated by the $\hdelta$-action completely breaks the bose-fermi cancellation 
in the untwisted Hilbert space.  
Moreover, any space-time supercharges\footnote
{In this paper, 
we shall regard the space-time supercharges as the operators acting consistently on the whole Hilbert space of 
string states and made up of the local perturbative degrees of freedom on the world-sheet. 
We use the term `non-SUSY string vacuum' with the meaning of absence of supercharges under this definition.  
It is beyond our scope whether non-perturbative supercharges could exist.    
In addition, it may be possible that one would gain some `supercharges' after truncating massive degrees of freedom 
in the approximation by low energy effective field theories.  
} 
cannot be constructed even if incorporating the degrees of freedom in the twisted sectors, 
so far as we assume the chiral forms of supercharges, namely, the integrals of conserved world-sheet current 
`$\dsp \cQ^{\al} = \oint d\bar{z}\, \cJ_R^{\al}(\bar{z})$', 
as was addressed in \cite{AS}. 
At this point it is crucial that the relevant twisted sectors are associated to  the {\em left-moving\/} operator $\delta_L$, whereas 
the possible supercharges should originate from the {\em right-moving\/} degrees of freedom.   

In the end, it is enough to ask whether or not the total partition function that contains all the twisted sectors vanish. 
We will clarify the `criterion' to this aim, and present a classification of the relevant orbifold models leading to   
the string vacua with the desired properties. 
In some cases, the desired vacua can be constructed in the manner similar  to those given in 
\cite{AS} for the type II string, in which the orbifold groups contain two generators equipped with some discrete torsions 
\cite{Vafa:1986wx,Vafa:1994rv,Gaberdiel:2000fe}. 
On the other hand, we also find out the  simpler models which  are just realized
as the asymmetric orbifolds of cyclic groups with only one generator, in contrast to the type II string cases.


~



\section{Preliminaries}
\label{sec:pre}

We begin with making  a very brief review of the heterotic string vacua 
including the Gepner models for the $\mbox{CY}_3$-compactifications, and 
prepare the notations to be used in the main section. 

~


\subsection{Heterotic String Vacua of Gepner Models}
\label{Gepner}

The Gepner model \cite{Gepner} describing some  
${\rm CY}_3$ is defined as 
the superconformal system;
\begin{eqnarray}
 && \left\lb \cM_{k_1}\otimes \cdots  \otimes
\cM_{k_r}\right\rb\left|_{\bsz_N\msc{-orbifold}}
\right. , ~~~ \sum_{i=1}^r \frac{k_i}{k_i+2}=3,
\label{Gepner CY3}
\end{eqnarray}
where $\cM_k$ denotes the $\cN=2$ minimal model of level $k$
($\dsp \hat{c}\equiv \frac{c}{3}= \frac{k}{k+2}$). 
We set
\begin{eqnarray}
 N := \mbox{L.C.M.} \, \{k_i+2~;~i=1,\ldots,r\}.
 \label{def N}
\end{eqnarray}
(`L.C.M.' means the least common multiplier.)
To describe the building blocks of torus partition function, 
we start with the  simple products of the characters of $\cN=2$ minimal model  \cite{Dobrev,RY1} in the NS-sector\footnote{We summarize the explicit character formulas as well as the conventions of theta functions in appendix A.
We set $q:= e^{2\pi i \tau}$, $y:= e^{2\pi i z}$ through this paper.};  
\begin{align}
& F^{(\sNS)}_I (\tau,z) := \prod_{i=1}^r \, \ch{(\sNS)}{\ell_i, m_i}(\tau,z),
\hspace{1cm} \left( I \equiv \left\{(\ell_i, m_i) \right\}, ~  ~ \ell_i+m_i \in 2\bz, \any i \right).
\label{FNS 0}
\end{align}
The ones for other spin structures are defined  by acting the half spectral flows $\dsp z\, \mapsto \, z+ \frac{r}{2} \tau + \frac{s}{2}$
($r,s \in \bz_2$):
\begin{align}
F^{(\stNS)}_I (\tau,z) & := F^{(\sNS)}_I \left(\tau,z+\frac{1}{2} \right),
\label{FtNS 0}
\\
F^{(\sR)}_I (\tau,z) & := q^{\frac{\hc}{8}} y^{\frac{\hc}{2}} \, F^{(\sNS)}_I\left(\tau, z+ \frac{\tau}{2} \right),
\label{FR 0}
\\
F^{(\stR)}_I (\tau,z) & := q^{\frac{\hc}{8}} y^{\frac{\hc}{2}} \, F^{(\sNS)}_I\left(\tau, z+ \frac{\tau+1}{2} \right),
\label{FtR 0}
\end{align}
where we set $\hc=3$. 
Note that the label $I \equiv \left\{(\ell_i, m_i)\right\}$ of the building blocks 
(and the spectral flow orbits introduced below) expresses the quantum numbers for the NS-sector 
{\em even for\/} 
$F^{(\sR)}_I$ and $F^{(\stR)}_I$. 

To construct the Gepner models, 
we need to  make the chiral $\bz_N\times \bz_N$ orbifolding by 
$g_L \equiv e^{2\pi i J^{\msc{tot}}_0}$ and $g_R \equiv e^{2\pi i \tJ^{\msc{tot}}_0}$, where 
$J^{\msc{tot}}$ ($\tJ^{\msc{tot}}$) expresses the  total $\cN=2$ $U(1)$-current in the left (right) mover acting over 
$\dsp \otimes_i \cM_{k_i}$. 
Recall that the zero-mode $J_0^{\msc{tot}}$ takes the eigen-values in $\dsp \frac{1}{N} \bz$ for the NS sector.
The chiral $\bz_N$-orbifolding (in the left mover) is represented in a way respecting the modular covariance
by considering the `spectral flow orbits' \cite{EOTY} defined as follows:
\begin{align}
\cF^{(\sNS)}_I (\tau,z) & := \frac{1}{N} \sum_{a,b\in \bz_N} \, q^{\frac{\hc}{2}a^2} y^{\hc a}F^{(\sNS)}_I \left(\tau,z+a\tau+b\right),
\label{cFNS}
\\
\cF^{(\stNS)}_{I} (\tau,z) &:= \cF^{(\sNS)}_{I} \left(\tau,z+\frac{1}{2}\right)
 \nn
 & \equiv \frac{1}{N} \sum_{a,b\in \bz_N} \, (-1)^{\hc a} q^{\frac{\hc}{2}a^2} y^{\hc a}F^{(\stNS)}_I \left(\tau,z+a\tau+b\right),
\label{cFtNS}
\\
\cF^{(\sR)}_{I} (\tau,z) &:= q^{\frac{\hc}{8}} y^{\frac{\hc}{2}}\cF^{(\sNS)}_{I} \left(\tau,z+\frac{\tau}{2}\right)
 \nn
 & \equiv \frac{1}{N} \sum_{a,b\in \bz_N} \, (-1)^{\hc b} q^{\frac{\hc}{2}a^2} y^{\hc a}F^{(\sR)}_I \left(\tau,z+a\tau+b\right),
\label{cFR}
\\
\cF^{(\stR)}_{I} (\tau,z) &:= q^{\frac{\hc}{8}} y^{\frac{\hc}{2}}\cF^{(\sNS)}_{I} \left(\tau,z+\frac{\tau+1}{2}\right)
 \nn
 & \equiv \frac{1}{N} \sum_{a,b\in \bz_N} \,  (-1)^{\hc (a+b)} q^{\frac{\hc}{2}a^2} y^{\hc a}F^{(\stR)}_I \left(\tau,z+a\tau+b\right).
\label{cFtR}
\end{align}
We also use the abbreviated notation; $\cF^{(\sigma)}_I(\tau) \equiv \cF^{(\sigma)}_I(\tau,0)$.
See Appendix B for the explicit forms of $\cF^{(\sigma)}_I(\tau,z)$ written
in terms of the $\cN=2$ minimal characters. 



~

Now, let us focus on the heterotic string. 
We take the convention;
$$
\mbox{left-mover} \, : \, \mbox{26D bosonic}, ~~~ \mbox{right-mover} \, : \,\mbox{10D super}.
$$
Assuming  the standard embedding of spin connection, 
the $SO(32)$ heterotic string vacuum compactified on 
$\mbox{CY}_3$
is described by the following modular invariant partition function;
\begin{align}
Z_{\msc{$SO(32)$ Het}}(\tau) & = \left(\frac{1}{\sqrt{\tau_2} \left|\eta\right|^2}\right)^2 \cdot 
\frac{1}{2 N} \, \sum_{\sigma_L, \sigma_R}\, 
\ep(\sigma_R) 
\left( \frac{\th_{[\sigma_L]}}{\eta}\right)^{13}
\overline{\left(\frac{\th_{[\sigma_R]}}{\eta}\right)}
\nn
& \hspace{2cm}
\times \sum_{I_L,I_R}\, N_{I_L,I_R} \cF^{(\sigma_L)}_{I_L}(\tau)
\overline{\cF^{(\sigma_R)}_{I_R}(\tau)}.
\label{SUSY Het Gepner SO(32)}
\end{align}
To avoid complexities, we shall assume the modular invariant coefficient $N_{I_L, I_R}$ to be diagonal through this paper:    
\begin{equation}
N_{I_L, I_R} \equiv \prod_{i=1}^r\, \frac{1}{2} \delta_{\ell_{i, L}, \ell_{i, R}} \delta_{m_{i,L}, m_{i, R}}, 
\hspace{1cm} \left(I_L \equiv \left\{ (\ell_{i, L}, m_{i, L} ) \right\}, ~~~ I_R \equiv \left\{ (\ell_{i, R}, m_{i, R} ) \right\} \right).
\label{mod inv N}
\end{equation}
Here the summations of $\sigma_L$, $\sigma_R$ are taken over the chiral spin structures.
We also set $\ep(\NS)= - \ep(\tNS) = - \ep(\R) = 1$ in the standard fashion, and $\th_{[\sNS]} \equiv \th_3(\tau,0)$, $\th_{[\stNS]} \equiv \th_4(\tau,0)$, 
$\th_{[\sR]} \equiv \th_2(\tau,0)$, $\left(\th_{[\stR]} \equiv - i \th_1(\tau,0) \equiv 0 \right)$ to describe the free fermion 
contributions including the $SO(32)$-sector.

The $E_8\times E_8$ heterotic string vacuum is likewise described 
as 
\begin{align}
Z_{\msc{$E_8\times E_8$ Het}}(\tau) & = \left(\frac{1}{\sqrt{\tau_2} \left|\eta\right|^2}\right)^2 \cdot 
\frac{1}{2 N} \, \sum_{\sigma_L, \sigma_R}\, 
\ep(\sigma_R) 
\left( \frac{\th_{[\sigma_L]}}{\eta}\right)^5 \chi_0^{E_8}
\overline{\left(\frac{\th_{[\sigma_R]}}{\eta}\right)}
\nn
& \hspace{2cm}
\times \sum_{I_L,I_R}\, N_{I_L,I_R} \cF^{(\sigma_L)}_{I_L}(\tau)
\overline{\cF^{(\sigma_R)}_{I_R}(\tau)},
\label{SUSY Het Gepner E8}
\end{align}
where $\chi_0^{E_8}(\tau) $ denotes the character of basic representation of affine $E_8$, written explicitly as 
\begin{equation}
\chi_0^{E_8}(\tau) \equiv \frac{1}{2} \left[\left(\frac{\th_3}{\eta}\right)^8 + \left(\frac{\th_4}{\eta}\right)^8
+\left(\frac{\th_2}{\eta}\right)^8 \right] (\tau).
\end{equation}


~


\section{Constructions of Non-SUSY Heterotic String Vacua}

In this section we present our main analysis. 
Namely, we discuss how we can construct the non-SUSY string vacua with 
the vanishing cosmological constant at one-loop (or the vanishing torus partition function)
based on the heterotic string compactified on $\mbox{CY}_3$ given in \eqn{SUSY Het Gepner SO(32)} and \eqn{SUSY Het Gepner E8}.
We start with specifying the relevant orbifold action.

~


\subsection{Orbifold Actions}

Let us fix a subsystem of the minimal models $\otimes_{i\in S}\, \cM_{k_i} $, 
$S \subset \{ 1, 2, \ldots, r \}$, on which the orbifold operators non-trivially 
act. We set
\begin{equation}
N' := \mbox{L.C.M.}\, \left\{k_i+2 \, :\, i \in S \right\} .
\end{equation}
The total central charge of the subsystem $S$ is written in the form;
\begin{equation}
\hc_S \left(\equiv \sum_{i \in S}\, \frac{k_i}{k_i+2}\right) 
= \left\{ 
\begin{array}{ll}
\dsp \frac{2 M}{N'}, 
~~
\ex M \in \bz.
& ~~ \left(N' \in 2\bz \right) 
\\
\dsp \frac{M'}{N'}, 
~~
\ex M' \in \bz,
& ~~ \left( N' \in 2\bz+1 \right)
\end{array} 
\right.
\label{hc S}
\end{equation}
We  fix a positive integer $L$ dividing $N'$,
and set 
\begin{equation}
4 K := \mbox{L.C.M}\, \left\{ N'/L, ~ 4 \right\},
\label{def K}
\end{equation}
for the later convenience.
We will soon define the orbifold action $\hdelta$ that satisfies  $\hdelta^{4K} = {\bm 1}$ on the untwisted sector.
We also define $S_1 \subset S$ by 
\begin{equation}
S_1 := \left\{ i \in S \, :\, \frac{N'}{k_i+2} \in 2\bz+1 \right\}.
\label{def S_1}
\end{equation} 
Note that $S_1 \neq \phi$ since  
$N'$ is the L.C.M. of $\{ k_i+2\}_{i\in S}$. 


For the $SO(32)$ ($E_8\times E_8$) heterotic string, we have the $SO(26)$ ($SO(10) \times E_8$) symmetry 
after making the standard embedding of spin connection. 
We will adopt the relevant orbifold action as a cyclic subgroup of $U(1)^s$ or $U(1)^{s_1} \times U(1)^{s_2}$
given as 
%
%
\begin{equation}
U(1)^s \times SO(26-2s) \subset SO(26),
\label{U(1)s}
\end{equation}
for the $SO(32)$-case, and 
%
%
\begin{equation}
\left[U(1)^{s_1} \times SO(10-2s_1)\right] 
\times \left[U(1)^{s_2} \times SO(16-2s_2) \right] 
\subset SO(10) \times E_8,
\label{U(1)s1s2}
\end{equation}
for the $E_8\times E_8$-case.


~

Now, let us specify the relevant orbifold action.
For the cases of $SO(32)$-heterotic string, 
we define 
\begin{equation}
\hdelta := (-1)^{F_R}\, \otimes \delta_{L}, 
\hspace{1cm}
\delta_L := 
e^{2\pi i L \sum_{i \in S}\, J_{L, 0}^{(i)}}\,
e^{2\pi i \frac{1}{2} \sum_{j=1}^s \, K_{L, 0}^{(j)}},
\label{def hdelta}
\end{equation}
where $J_L^{(i)}$ is the left-moving $U(1)$-current in $\cM_{k_i}$, $i\in S$, and 
$K_L^{(j)}$ are those for the $U(1)^s$-factor in \eqn{U(1)s}. 
In other words, $\delta_L$ acts on the left-moving characters of $\cM_{k_i}$, $i\in S$ as 
the integral spectral flow $\dsp z\, \mapsto \, z + L (\al \tau + \beta)$;
\begin{align}
& \delta_{L, (\al,\beta)} \cdot \ch{(\sigma)}{\ell_i ,m_{i}}(\tau,z) := 
q^{\frac{k_i}{2(k_i+2)}L^2 \al^2} y^{\frac{k_i}{k_i+2} L \al} 
e^{2\pi i \frac{k_i}{2(k_i+2)} L^2 \al\beta}\,
\ch{(\sigma)}{\ell_i,m_i}\left(\tau,z+L \left(\al \tau + \beta\right)\right),
\nn
& \hspace{10cm}
(\al,\beta)\in \bz_{N'/L} \times \bz_{N'/L} ,
\label{deltaL ch}
\end{align}
which yields the modular covariant actions on the spectral flow orbits 
$\cF^{(\sigma_L)}_{I_L}(\tau)$. 
We summarize the explicit forms of  `$\delta_{L, (\al,\beta)} \cdot \cF^{(\sigma_L)}_{I_L}(\tau)$'
in Appendix B.
$F_R$ denotes the space-time fermion number of the right-mover. Namely, 
the operator $(-1)^{F_R}$ acts as the sign-flip of the right-moving R-sector.  

On the other hand, $\delta_L$ acts on the Jacobi's theta functions associated to the $U(1)^s$-factor as follows;
\begin{align}
& \delta_{L, (\al,\beta)} \cdot \th_i(\tau,z) 
\left(
\equiv \th_{i,(\al,\beta)}(\tau,z)
\right)
:= 
q^{\frac{\al^2}{8}} y^{\frac{\al}{2}} e^{2\pi i \frac{\al\beta}{8}}\,
\th_i\left(\tau, z+ \frac{\al \tau+\beta}{2} \right).
\hspace{1cm} \left(i=3,4,2 \right)
\label{deltaL th}
\end{align}
Here the inclusion of  phase factor $e^{2\pi i \frac{\al\beta}{8}}$ is necessary for the modular covariance as in the minimal sector 
\eqn{deltaL ch}.
The explicit forms of 
\eqn{deltaL th}
are also summarized in Appendix B.


We similarly define the orbifold action $\hdelta$ 
in the $E_8\times E_8$-case,
in which $\delta_L$ acts on
$$
\bigotimes_{i\in S} \cM_{k_i} \bigotimes U(1)^{s_1} \otimes U(1)^{s_2},
$$
in the same way as \eqn{def hdelta}.


~

Since the $\hdelta$-orbifold action  is defined so as to respect the modular covariance, 
it is easy to write down the modular invariant partition functions of our asymmetric orbifolds. 
For example, for the $SO(32)$ heterotic string and in the cases of $K s \in 2\bz+1$, 
the $\hdelta$-orbifold is found to be order $8K$,  and the modular invariant parttion function is 
written as 
\begin{align}
Z_{\hdelta\msc{-orb}}(\tau) & = \left(\frac{1}{\sqrt{\tau_2} \left|\eta\right|^2}\right)^2 \cdot 
\frac{1}{2 N} \, \sum_{\sigma_L, \sigma_R}\, 
\frac{1}{8K}\sum_{\al, \beta \in \bz_{8K}}\, 
\ep(\sigma_R; \al,\beta) 
\left( \frac{\th_{[\sigma_L], (\al,\beta)}}{\eta}\right)^{s}
\left( \frac{\th_{[\sigma_L]}}{\eta}\right)^{13-s}
\overline{\left(\frac{\th_{[\sigma_R]}}{\eta}\right)}
\nn
& \hspace{1cm}
\times \sum_{I_L,I_R}\, N_{I_L,I_R} 
\left[ \delta_{L, (\al,\beta)} \cdot\cF^{(\sigma_L)}_{I_L}(\tau) \right]
\overline{\cF^{(\sigma_R)}_{I_R}(\tau)}.
\label{Non-SUSY Het Gepner SO(32)}
\end{align}
Here, we set 
\begin{align}
&
\ep(\NS; \al,\beta)
:=
\left\{
\begin{array}{ll}
-1 & ~~ (\al,\beta \in 2\bz+1)
\\
1 & ~~ (\mbox{otherwise})
\end{array}
\right.
\nn
&
\ep(\tNS; \al,\beta)
:=
\left\{
\begin{array}{ll}
1 & ~~ (\al \in 2\bz+1, ~ \beta\in 2\bz)
\\
-1 & ~~ (\mbox{otherwise})
\end{array}
\right.
\nn
&
\ep(\R; \al,\beta)
:=
\left\{
\begin{array}{ll}
1 & ~~ (\al \in 2\bz, ~ \beta\in 2\bz+1)
\\
-1 & ~~ (\mbox{otherwise})
\end{array}
\right.
\label{def ep al beta}
\end{align}
which originate from the GSO phases $\ep(\sigma_R)$ modified by the $(-1)^{F_R}$-actions included in \eqn{def hdelta}.  
Also, 
we again made use of the abbreviated notations  
$\th_{[\sNS],(\al,\beta)}(\tau) \equiv \th_{3,(\al,\beta)}(\tau) \equiv \th_{3,(\al,\beta)}(\tau,0)$, and so on.

The modular invariants in other cases are similarly obtained.

~


\subsection{Criterion for the Desired Models}

At this stage
let us clarify the `criterion' to search for the heterotic string vacua with 
the desired properties. 
To this end, we denote the contributions to the torus partition function 
from the each twisted sector as `$Z_{(\al,\beta)}(\tau)$', ($\al,\beta \in \bz_{4K}$).
That is, we define 
\begin{align}
Z_{(\al,\beta)}(\tau) \equiv \tr_{\hdelta^{\al}\msc{-twisted}} \, 
\left[ \hdelta^\beta \, q^{L_0-\frac{\hc}{8}} \overline{q^{\tL_0-\frac{\hc}{8}}} \right],
\label{def Z al beta}
\end{align}
for the convenience. 
By our definition of the orbifold action $\hdelta$ presented above,
the building blocks $Z_{(\al,\beta)}(\tau)$ covariantly behave under the modular transformations;
\begin{align}
&& Z_{(\al,\beta)}\left(-\frac{1}{\tau} \right) = Z_{(\beta, -\al)} (\tau), 
\hspace{1cm}
Z_{(\al,\beta)}(\tau+1) = Z_{(\al,\al+\beta)}(\tau). 
\label{Z mod cov}
\end{align}

We require the following conditions;
\begin{itemize}
\item For the `even sectors' $\any \al, \beta \in 2\bz$, each building block $Z_{(\al,\beta)}(\tau)$ separately vanishes;
\begin{align}
Z_{(\al,\beta)}(\tau) \equiv 0,
\label{even sec cond}
\end{align}


\item
The partition function for the untwisted sector does not vanish;
\begin{align}
Z_{0} (\tau) \equiv \frac{1}{4 K }\sum_{\beta \in \bz_{4K}}\, Z_{(0,\beta)} \not\equiv 0,
\label{ut sec cond}
\end{align}


\item 
For all the twisted sectors of $\hdelta^{\al}$ with $\al \in 2\bz+1$, we require
\begin{align}
\sum_{\beta'}\, Z_{(\al,2\beta')}(\tau) \equiv 0.
\hspace{1cm} 
\left(\any \al \in 2\bz+1 \right)
\label{odd sec cond}
\end{align}

\end{itemize}

~

Note that, 
\eqn{odd sec cond} just implies
\begin{align}
\sum_{\al \in 2\bz+1 \, \msc{or} \, \beta \in 2\bz+1} \, Z_{(\al,\beta)}(\tau) \equiv 0,
\label{odd sec vanish}
\end{align}
due to the modular covariance \eqn{Z mod cov}.
Thus, combining it together with the requirement \eqn{even sec cond}, we can conclude that 
the total partition function should vanish.

We also note that, in this situation, the bose-fermi cancellation  can only  occur 
among the {\em different\/} twisted sectors because of the condition \eqn{ut sec cond}. 
On the other hand, the possible space-time supercharges should be of the form such as 
$\dsp \cQ^{\al} = \oint d\bar{z}\, \cJ_R^{\al}(\bar{z})$, 
which is consistent with the conservation on the world-sheet.
However, any operators of this form cannot induce the expected bose-fermi cancellation, because
the relevant twisted sectors are associated to the {\em left-moving\/} operator $\delta_L$.
In this way, we conclude that we do not have any space-time supercharges as the operators 
consistently acting on the whole Hilbert space and conserved on the world-sheet.
This is the reason why we claim that the heterotic string vacua 
that satisfy the above requirements are non-supersymmetric ones.


~


\subsection{Classification of the Models}

We here study 
the aspects of  orbifolds of heterotic string vacua 
\eqn{SUSY Het Gepner SO(32)} and \eqn{SUSY Het Gepner E8}
by the cyclic actions of $\hdelta$ given in \eqn{def hdelta}.
We classify the models according to the positive integer $N'/L$.

First of all, we note 
\begin{align}
Z_{(\al,\beta)}(\tau) \equiv 0, 
\hspace{1cm}
\any \al, \beta \in 2\bz,
\label{Z even}
\end{align}
for all the cases we will discuss below, since 
$\hdelta^2$ obviously preserves the space-time supercharges.   

One can also readily confirm that, for the untwisted sector $\al=0$, 
\begin{equation}
Z_0(\tau) \left( \equiv \frac{1}{4K} \sum_{\beta \in \bz_{4K}}\, Z_{(0,\beta)} (\tau) \right) \neq 0, 
\end{equation}
in all the cases.

~


Now, let us describe the classification:

~

\noindent
{\bf 1. $N'/L \equiv 0$ $(\mod\, 4)$ :}

In this case we have $\dsp \frac{N'}{L} = 4K$. 

We first focus on the $SO(32)$-case. 
The crucial point is as follows:
For the product of left-moving minimal characters $\dsp \prod_{i\in S} \ch{(\sigma)}{\ell_{i,L}, m_{i,L}-2n_L}(\tau)$ 
($n_L$ is the spectral flow momentum) as well as 
the each `free fermion factor' $\dsp \left(\frac{\th_j}{\eta}\right)^{13}$, 
the orbifold action $\hdelta$ picks up the next phase factor;
\begin{align}
& \exp \left[2\pi i \left\{ \sum_{i\in S}\, \frac{d_i(m_{i,L}-2n_L)}{N'} L \beta + \frac{M}{N'} L^2 \al \beta + \frac{1}{8} c_j s \al \beta\right\}\right]
\nn
& \equiv \exp \left[2 \pi i \frac{\beta}{4K} \left\{ \sum_{i\in S}  d_i (m_{i,L}-2n_L) +  M L \al + \frac{1}{2} K c_j s \al \right\}\right],
\label{phase even 1}
\end{align}
for the $(\al,\beta)$-twisted sector with $\beta \in 2\bz$.
Here we set $\dsp d_i := \frac{N'}{k_i+2}$ for $\any i \in S$, and $c_j = 1$ ($c_j=-1$) for $j=3,4$ (2). 
We also note that, when $\beta \in 2\bz+1$,  the similar phase factor is gained, while the $\dsp \left(\frac{\th_3}{\eta}\right)^s$ factor 
is exchanged with 
$\dsp \left(\frac{\th_4}{\eta}\right)^s$.

Fixing the value  $\al \in 2\bz+1$, 
let us evaluate the summation 
$\dsp 
\sum_{\beta'} \, Z_{(\al,2\beta')}(\tau).
$
It acts as the projection 
imposing
\begin{align}
\sum_{i\in S} \, d_i(m_{i, L}-2n) + ML \al + \frac{1}{2} K c_j s \al \equiv 0 ~ (\mod\, 2K).
\label{constraint 1}
\end{align}


The arguments are almost the same for the $E_8\times E_8$-case. 
We only have to replace 
the term $\dsp \frac{1}{2}K c_j s$ in \eqn{phase even 1}
with $\dsp \frac{1}{2}K \left(c^{(1)}_j s_1+ c^{(2)}_k s_2\right)$
for the factor 
$
\dsp \left(\frac{\th_j}{\eta}\right)^{s_1} \left(\frac{\th_k}{\eta}\right)^{s_2},
$
where $c^{(r)}_j= \pm 1$ is defined in the same way as above.  
Thus, the constraint \eqn{constraint 1} is replaced with 
\begin{align}
\sum_{i\in S} \, d_i(m_{i, L}-2n) + ML \al + \frac{1}{2} K \left(c^{(1)}_j s_1 + c^{(2)}_k s_2 \right) \al \equiv 0 ~ (\mod\, 2K).
\label{constraint 1'}
\end{align}

~

Consequently, we obtain the next classification;
\begin{itemize}
\item $K s \in 4\bz $ [ $SO(32)$ ]\\
 $K s_1, K s_2 \in 4\bz$ or $K s_1, K s_2 \in 4\bz+2$ [ $E_8 \times E_8$ ] :

In these cases the aspects are almost parallel to that of \cite{AS}.
The constraint \eqn{constraint 1} or \eqn{constraint 1'}  implies 
\begin{equation}
\sum_{i \in S_1}\, m_{i,L} + LM \equiv 0 ~ (\mod\, 2),
\label{constraint 1-1}
\end{equation}
where $S_1$ has been defined in \eqn{def S_1}, that is,  
\begin{equation}
S_1 \equiv  \left\{i\in S~:~ d_i \in 2\bz+1\right\}.
\nonumber
\end{equation}
We then find 
$$
\sum_{\beta'} \, Z_{(\al,2\beta')}(\tau) \neq 0, 
$$
since we generically possess many states satisfying the condition \eqn{constraint 1-1}. 
This means that $\hdelta$-orbifolding cannot satisfy \eqn{odd sec cond} by itself.

However, as was shown in \cite{AS}\footnote
   {In comparison with \cite{AS}, the roles of left and right movers have been exchanged here.}, we can make it possible 
by further introducing  the $\bz_2$-orbifold action $\hgamma$, which commutes with $\hdelta$;
\begin{equation}
\hgamma := \left\{ 
\begin{array}{ll}
 (-1)^{F_L} \otimes \gamma_R
& ~~ (\# S_1 \in 2\bz+1), \\
\gamma_R 
& ~~ (\# S_1 \in 2\bz) ,
\end{array}
\right.
\label{hgamma}
\hspace{1cm}
\gamma_R := \prod_{i\in S_1}\, (-1)^{\ell_{i, R}}.
\end{equation}
on the right-moving minimal characters $\dsp \overline{\prod_{i\in S} \ch{(\sigma)}{\ell_{i,R}, m_{i,R}-2n_R}(\tau)}$, 
and $(-1)^{F_L}$ denotes the sign-flip of the left-moving R-sector.  
We shall also introduce  the discrete torsion 
\cite{Vafa:1986wx,Vafa:1994rv,Gaberdiel:2000fe}
with respect to the $\hgamma$ and $\hdelta$-actions;
\begin{equation}
\xi \left( a, \al \, ; b ,\beta \right) := (-1)^{\left(LM-1\right) \left(a\beta - b \al\right)},
\hspace{1cm} \left(a, b \in \bz_2, ~ \al, \beta \in \bz_{4K} \right),
\label{dtorsion S}
\end{equation}
where $a,b$ label the spatial and temporal twistings by $\hgamma$, while $\al,\beta$ are those associated to $\hdelta$ as above.  
Then, for any fixed $ \al \in 2\bz+1$, we readily obtain 
\begin{align}
\sum_{\beta'} \, \left. Z_{(\al,2\beta')}(\tau)\right|_{\hgamma-\msc{orbifold}} & \equiv  
\frac{1}{2} \sum_{\beta'}\, \sum_{a,b\in \bz_2}\,  \xi \left( a, \al \, ; b ,2\beta' \right)\, Z_{(a, \al ; \, b, 2\beta')} (\tau) =0.
\end{align}
That is, the criterion \eqn{odd sec cond} is satisfied when combining the orbifolding by $\hgamma$ and $\hdelta$.
See \cite{AS} for more detailed arguments.



\item $K s \in 4\bz+2 $ [$SO(32)$] \\
$K s_1 \in 4\bz$, $K s_2 \in 4\bz+2$, or $K s_1 \in 4\bz+2$, $K s_2 \in 4\bz$ [$E_8\times E_8$] :

In these cases, the constraint \eqn{constraint 1} or \eqn{constraint 1'} yields 
\begin{equation}
\sum_{i \in S_1}\, m_i + LM +1 \equiv 0 ~ (\mod\, 2),
\label{constraint 1-2}
\end{equation}
in place of \eqn{constraint 1-1}.
Therefore, we can make the criterion \eqn{odd sec cond} to be satisfied by taking again the $\hdelta$ and $\hgamma$ orbifolds
but with the different discrete torsion 
 \begin{equation}
\xi' \left( a, \al \, ; b ,\beta \right) := (-1)^{LM \left(a\beta - b \al\right)},
\hspace{1cm} \left(a, b \in \bz_2, ~ \al, \beta \in \bz_{4K} \right).
\label{dtorsion S'}
\end{equation}


\item $K s \in 2\bz+1 $ [$SO(32)$] \\
$K(s_1+s_2) \in 2\bz+1$ [$E_8\times E_8$] :

In these cases, any states cannot satisfy 
the condition \eqn{constraint 1}, 
and thus the criterion \eqn{odd sec cond} is trivially achieved
by only making the $\hdelta$-orbifolding.


\item $K s_1 , K s_2 \in 2\bz+1$ [$E_8\times E_8$] :

In these remaining cases, both of \eqn{constraint 1-1} and \eqn{constraint 1-2} are possible, depending on which theta function 
factors $(\th_j)^{s_1} (\th_k)^{s_2}$ the operator $\hdelta$ acts. 
Thus, \eqn{odd sec cond} cannot be satisfied even if incorporating the $\hgamma$-orbifolding. 
We conclude that the string vacua with the desired properties are not constructed in these cases.  

\end{itemize}


~

\noindent
{\bf 2. $N'/L \not\equiv 0$ $(\mod\, 4)$ : }

In this case, 
$\dsp \frac{N'}{L} \in 4\bz+2$ or $\dsp \frac{N'}{L} \in 2\bz+1$, and 
$K \in 2\bz+1$ for the both cases.

Again, we first consider the $SO(32)$ case.
In the case of $\dsp \frac{N'}{L} \in 4\bz+2$, the $\hdelta$ picks up the phase factor;
\begin{align}
& \exp \left[2\pi i \left\{ \sum_{i\in S}\, \frac{d_i(m_i-2n)}{N'} L \beta + \frac{M}{N'} L^2 \al \beta + \frac{1}{8}s c_j \al \beta\right\}\right]
\nn
& \equiv \exp \left[2 \pi i \frac{\beta}{4K} \left\{ \sum_{i\in S} 2 d_i (m_i-2n) + 2 M L \al + \frac{1}{2} K s c_j \al \right\}\right],
\label{phase even 2}
\end{align}
instead of \eqn{phase even 1}.
In the case of $\dsp \frac{N'}{L} \in 2\bz+1$, we similarly obtain
\begin{align}
& \exp \left[2\pi i \left\{ \sum_{i\in S}\, \frac{d_i(m_i-2n)}{N'} L \beta + \frac{M'}{2N'} L^2 \al \beta + \frac{1}{8}s c_j \al \beta\right\}\right]
\nn
& \equiv \exp \left[2 \pi i \frac{\beta}{4K} \left\{ \sum_{i\in S} 4 d_i (m_i-2n) + 2 M' L \al + \frac{1}{2} K s c_j \al \right\}\right].
\label{phase odd}
\end{align}


In the case of $E_8\times E_8$, 
the term $\dsp \frac{1}{2}K c_j s$ in \eqn{phase even 2} and \eqn{phase odd} is again replaced 
with \\ $\dsp \frac{1}{2}K \left(c^{(1)}_j s_1+ c^{(2)}_k s_2\right)$.


Combining all the things, we obtain the next classifications;
\begin{itemize}
\item $s \not\in 4\bz$ [$SO(32)$] \\
$s_1+s_2 \in 2\bz+1$, ~ $s_1 \in 4\bz, \, s_2 \in 4\bz+2$, or $s_1 \in 4\bz+2, \,  s_2 \in 4\bz$ [$E_8\times E_8$] :

In all these cases we simply obtain 
$$
\sum_{\beta'}\, Z_{(\al,2\beta')} = 0, \hspace{1cm} \left(\any \al \in 2\bz+1\right), 
$$
because $\dsp \frac{1}{2} K s c_j \al$  (or $\dsp  \frac{1}{2}K \left(c^{(1)}_j s_1+ c^{(2)}_k s_2 \right)\al $ for $E_8\times E_8$)
takes values in $\dsp \bz+\frac{1}{2} $ or $2\bz+1$, and thus the phase factors \eqn{phase even 2}, \eqn{phase odd} never cancel out.   
Hence, the criterion \eqn{odd sec cond} is again achieved
by making only the $\hdelta$-orbifolding.

\item otherwise : 

In the remaining cases, we have 
$
\dsp 
\sum_{\beta'}\, Z_{(\al,2\beta')} \neq 0.
$
Moreover, \eqn{odd sec cond} cannot be satisfied even if the $\hgamma$-orbifolding is incorporated with any discrete torsion.
The desired string vacua are not constructed in these cases.

\end{itemize}


~

To summarize, we have obtained the non-SUSY heterotic string vacua with the property $Z_{\msc{1-loop}}(\tau) \equiv 0$
based on the orbifolding by $\hdelta$ (and $\hgamma$ in some cases) as follows:


\begin{description}
\item[(1)] $Ks \in 2\bz+1$ $[SO(32)]$, \\
$K(s_1+s_2) \in 2\bz+1$ $[E_8\times E_8]$ :

The desired vacua can be constructed only by making the $\hdelta$-orbifolding. 
The order of orbifolding is $8K$ although $\hdelta^{4K} = \bm{1}$ if restricting on the untwisted Hilbert space.

\item[(2)] $Ks \in 4\bz+2$ and $N'/L \not\equiv 0 ~(\mod\, 4)$ $[SO(32)]$, \\
$K s_i \in 4\bz+2$, $K s_j \in 4\bz$ ($i\neq j$) and $N'/L \not\equiv 0 ~(\mod\, 4)$ $[E_8\times E_8]$ :

The wanted vacua are again constructed only by $\hdelta$-action as in the case {\bf (1)}. 
However, we obtain an order $4K$ orbifold in this case.

\item[(3)] $Ks \in 2\bz$ and $N'/L \equiv 0~ (\mod\, 4)$ $[SO(32)]$, \\
$K s_1, \, K s_2 \in 2\bz$ and $N'/L \equiv 0~ (\mod\, 4)$ $[E_8\times E_8]$ :

The wanted vacua are constructed as the $\bz_{4K} \times \bz_2$-orbifold defined by $\hdelta$ and $\hgamma$-actions 
with the next discrete torsion included ($a, b \in \bz_2 $ for $\hgamma$-twists, and $\al, \beta \in \bz_{4K} $ for $\hdelta$-twists);
\begin{equation}
\xi \left( a, \al \, ; b ,\beta \right) = 
\left\{
\begin{array}{ll}
\dsp (-1)^{\left(LM+\frac{1}{2} Ks-1\right) \left(a\beta - b \al\right)},
& ~~ [SO(32)]
\\
\dsp (-1)^{\left(LM+\frac{1}{2} K(s_1+s_2) -1\right) \left(a\beta - b \al\right)}.
& ~~ [E_8\times E_8]
\end{array}
\right.
\label{dtorsion S general}
\end{equation}

\end{description}


~

\section{Some Comments}

In this paper, as an extension of our previous work \cite{AS}, we have studied 
the construction of non-SUSY heterotic string vacua with the vanishing cosmological constant at one-loop,
based on the asymmetric orbifolding of the Gepner models. 
We would like to add a few comments:


\begin{itemize}
\item 
In the string vacua we constructed, we could not make up the space-time supercharges that 
are conserved   
on the world-sheet and consistently realizing the bose-fermi cancellation expected from the one-loop partition functions.
We would like to here emphasize that $Z_{\msc{one-loop}}(\tau) \equiv 0$ just implies the bose-fermi cancellation 
{\em under the free string limit. }
Therefore, even if they might induce some low-energy effective field theories with unbroken SUSY,
the absence of supercharges in the above sense should imply that 
they could not be  supersymmetric ones {\em when turning on the string interactions} described by 
general world-sheets with higher genera. 
It would be thus capable for them to generate small non-vanishing cosmological constants after incorporating the  
(perturbative or non-perturbative) stringy quantum corrections, 
although such analyses  still look very hard to carry out  
due to the complexities of spectra arising from various twisted sectors. 



\item 
When being motivated by the cosmological constant problem, it would be more desirable but much more non-trivial situations  
that we have 
the vanishing one-loop cosmological constant {\em without\/} the bose-fermi cancellation at each mass level
(in other words, 
$Z(\tau) \not\equiv 0$, but 
$
\dsp 
\La \equiv \int \frac{d^2\tau}{\tau_2^2} \, Z(\tau) =0 
$).
On the other hand, 
a characteristic feature of the string vacua given in the present paper (and those given in \cite{AS}) is 
that we have the bose-fermi cancellation among the different twisted sectors of the relevant orbifolding, 
as was emphasized several times.   
We would like to discuss elsewhere the possibility to realize such `desirable situations', 
at least, in some {\em point particle\/} theories 
with infinite mass spectra  (not necessarily, string theories), by implementing this feature
\cite{SS work in progress}.

\end{itemize}


~





\newpage


\appendix

\section*{Appendix A: ~ Summary of  Conventions}

\setcounter{equation}{0}
\def\theequation{A.\arabic{equation}}

~

We summarize the notations and conventions adopted in this paper. 
We set $q \equiv e^{2\pi i \tau}$, $y \equiv e^{2\pi i z}$.

~

\noindent
{\bf 1. Theta Functions} 
%
 \begin{align}
 & \dsp \th_1(\tau,z):=i\sum_{n=-\infty}^{\infty}(-1)^n q^{(n-1/2)^2/2} y^{n-1/2}
  \equiv  2 \sin(\pi z)q^{1/8}\prod_{m=1}^{\infty}
    (1-q^m)(1-yq^m)(1-y^{-1}q^m), \nn [-10pt]
   & \\[-5pt]
 & \dsp \th_2(\tau,z):=\sum_{n=-\infty}^{\infty} q^{(n-1/2)^2/2} y^{n-1/2}
  \equiv 2 \cos(\pi z)q^{1/8}\prod_{m=1}^{\infty}
    (1-q^m)(1+yq^m)(1+y^{-1}q^m), \\
 & \dsp \th_3(\tau,z):=\sum_{n=-\infty}^{\infty} q^{n^2/2} y^{n}
  \equiv \prod_{m=1}^{\infty}
    (1-q^m)(1+yq^{m-1/2})(1+y^{-1}q^{m-1/2}),  
\\
 &  \dsp \th_4(\tau,z):=\sum_{n=-\infty}^{\infty}(-1)^n q^{n^2/2} y^{n}
  \equiv \prod_{m=1}^{\infty}
    (1-q^m)(1-yq^{m-1/2})(1-y^{-1}q^{m-1/2}) . 
\\
& \Th{m}{k}(\tau,z):=\sum_{n=-\infty}^{\infty}
 q^{k(n+\frac{m}{2k})^2}y^{k(n+\frac{m}{2k})} ,
\\
&
\eta(\tau) := q^{1/24}\prod_{n=1}^{\infty}(1-q^n).
 \end{align}
 Here, we have set $q:= e^{2\pi i \tau}$, $y:=e^{2\pi i z}$  
 ($\any \tau \in \bh^+$, $\any z \in \bc$),
 and used abbreviations, $\th_i (\tau) \equiv \th_i(\tau, 0)$
 ($\th_1(\tau)\equiv 0$), 
$\Th{m}{k}(\tau) \equiv \Th{m}{k}(\tau,0)$.
%

~


\noindent
{\bf 2. Character Formulas for $\cN=2$ Minimal Model}

The character formulas of 
the level $k$ $\cN=2$ minimal model $(\hat{c}=k/(k+2))$ \cite{Dobrev,RY1}
are described 
as the branching functions of 
the Kazama-Suzuki coset \cite{KS} $\dsp \frac{SU(2)_k\times U(1)_2}{U(1)_{k+2}}$
defined by
\begin{eqnarray}
&& \chi_{\ell}^{(k)}(\tau,w)\Th{s}{2}(\tau,w-z)
=\sum_{\stackrel{m\in \bsz_{2(k+2)}}{\ell+m+s\in 2\bsz}} \chi_m^{\ell,s}
(\tau,z)\Th{m}{k+2}(\tau,w-2z/(k+2))~, \nn
&& \chi^{\ell,s}_m(\tau,z) \equiv  0~, ~~~ \mbox{for $\ell+m+s \in 2\bz+1$}~,
\label{branching minimal}
\end{eqnarray}
where $\chi_{\ell}^{(k)}(\tau,z)$ is the spin $\ell/2$ character of 
$SU(2)_k$;
\begin{eqnarray}
&&\chi^{(k)}_{\ell}(\tau, z) 
=\frac{\Th{\ell+1}{k+2}(\tau,z)-\Th{-\ell-1}{k+2}(\tau,z)}
                        {\Th{1}{2}(\tau,z)-\Th{-1}{2}(\tau,z)}
\equiv \sum_{m \in \bsz_{2k}}\, c^{(k)}_{\ell,m}(\tau)\Th{m}{k}(\tau,z)~.
\label{SU(2) character}
\end{eqnarray}
The branching function $\chi^{\ell,s}_m(\tau,z)$ 
is explicitly calculated as follows;
\begin{equation}
\chi_m^{\ell,s}(\tau,z)=\sum_{r\in \bsz_k}c^{(k)}_{\ell, m-s+4r}(\tau)
\Th{2m+(k+2)(-s+4r)}{2k(k+2)}(\tau,z/(k+2))~.
\end{equation}
Then, the character formulas of unitary representations 
are written as 
\begin{eqnarray}
&& \ch{(\sNS)}{\ell,m}(\tau,z) = \chi^{\ell,0}_m(\tau,z)
+\chi^{\ell,2}_m(\tau,z),
\nn
&& \ch{(\stNS)}{\ell,m}(\tau,z) = \chi^{\ell,0}_m(\tau,z)
-\chi^{\ell,2}_m(\tau,z)
\nn
&& \ch{(\sR)}{\ell,m}(\tau,z) = \chi^{\ell,1}_m(\tau,z)
+\chi^{\ell,3}_m(\tau,z) 
\nn
&& \ch{(\stR)}{\ell,m}(\tau,z) = \chi^{\ell,1}_m(\tau,z)
-\chi^{\ell,3}_m(\tau,z) .
\label{minimal character}
\end{eqnarray}

~


\section*{Appendix B: ~ Explicit Forms of Building Blocks and Their Orbifold Twistings}

\setcounter{equation}{0}
\def\theequation{B.\arabic{equation}}

~

In Appendix B, we summarize the explicit expressions of 
spectral flow orbits \eqn{cFNS}-\eqn{cFtR} playing the role of building blocks of relevant modular invariants.
We also describe the orbifold actions $\delta$, $\gamma$ on the spectral flow orbits, as well as 
the $\delta$-twistings on the theta function factor, denoted as `$\th_{i,(\al,\beta)}(\tau,z)$.'


We make use of the abbreviated index $I \equiv \left\{(\ell_i, m_i) \right\}$ ($\ell_i+m_i \in 2\bz$) again, and 
set 
\begin{equation}
Q(I) \equiv Q \left(\left\{(\ell_i, m_i) \right\}\right) := \sum_{i=1}^r \, \frac{m_i}{k_i+2} \left(\in \frac{1}{N}\bz\right),
\end{equation}
for the convenience. 
$\cF^{(\sigma)}_I(\tau,z) $ obviously vanishes  for $Q(I) \not\in \bz $ by the definitions \eqn{cFNS}-\eqn{cFtR}, 
and  
we obtain the following expressions in the case of $Q(I) \in \bz$,  
\begin{align}
\cF^{(\sNS)}_I (\tau,z) & = \sum_{n\in\bz_N}\, \prod_{i=1}^r\, \ch{(\sNS)}{\ell_i, m_i - 2n}(\tau,z)
\equiv \sum_{n\in\bz_N}\, F^{(\sNS)}_{s_n(I)}(\tau, z),
\label{cFNS 2}
\\
\cF^{(\stNS)}_I (\tau,z) & = (-1)^{Q(I)} \sum_{n\in\bz_N}\, (-1)^{\left(\hc+r\right) n} \prod_{i=1}^r\, \ch{(\stNS)}{\ell_i, m_i - 2n}(\tau,z)
\equiv \sum_{n\in\bz_N}\, (-1)^{\left(\hc+r\right) n} F^{(\stNS)}_{s_n(I)}(\tau, z),
\label{cFtNS 2}
\\
\cF^{(\sR)}_I (\tau,z) & = \sum_{n\in\bz_N}\, \prod_{i=1}^r\, \ch{(\sR)}{\ell_i, m_i - 2n-1 }(\tau,z)
\equiv \sum_{n\in\bz_N}\, F^{(\sR)}_{s_n(I)}(\tau, z),
\label{cFR 2}
\\
\cF^{(\stR)}_I (\tau,z) & = (-1)^{Q(I)+r} \sum_{n\in\bz_N}\, 
(-1)^{\left(\hc+r\right) n} \prod_{i=1}^r\, \ch{(\stR)}{\ell_i, m_i - 2n-1}(\tau,z)
\equiv \sum_{n\in\bz_N}\, (-1)^{\left(\hc+r\right) n} F^{(\stR)}_{s_n(I)}(\tau, z),
\label{cFtR 2}
\end{align}
where we introduced the notation
$$
s_n(I) := \left\{ (\ell_i, m_i -2n)\right\} ~~  
\left( \mbox{for} ~  I \equiv \left\{ (\ell_i, m_i)\right\}\right) .
$$

Then, the actions of $\delta$-twisting\footnote
  {Here we omit the subscripts `$L$' and `$R$' used in the main text.} for   
$\cF^{(\sigma)}_I(\tau)$ are expressed explicitly  as 
\begin{align}
\delta_{(\al,\beta)} \cdot \cF^{(\sigma)}_{\left\{(\ell_i, m_i) \right\}} (\tau)
& = \zeta_{\hc_S L}(\sigma; \al , \beta)\, e^{2\pi i \frac{L^2M}{N'} \al \beta} 
\sum_{n\in \bz_N}\, e^{2\pi i \sum_{i\in S}\, \frac{L \left(m_i-2n\right)}{k_i+2}\beta} \, 
F^{(\sigma)}_{\left\{(\ell_i, m''_i-2n)\right\}} (\tau)
\nn
& \equiv \zeta_{2LM/N'}(\sigma; \al , \beta)\, e^{2\pi i \frac{L^2M}{N'} \al \beta} 
\sum_{n\in \bz_N}\, e^{2\pi i \frac{L}{N'} \sum_{i\in S}\, d_i (m_i-2n)  \beta} \, 
F^{(\sigma)}_{\left\{(\ell_i, m''_i-2n )\right\}} (\tau),
\nn
& 
\hspace{8cm}
\left(d_i \equiv \frac{N'}{k_i+2} \right),
\label{delta cF S} 
\end{align}
where we introduced the phase factor
\begin{equation}
\zeta_{\kappa}(\NS; \al, \beta) =1, ~~\zeta_{\kappa}(\tNS; \al, \beta) =e^{i\pi \kappa\al}, ~~ 
\zeta_{\kappa}(\R; \al, \beta) = e^{i\pi \kappa \beta}, ~~
\zeta_{\kappa}(\tR; \al, \beta) = e^{i\pi \kappa(\al+\beta)}, 
\label{zeta kappa}
\end{equation}
and set
$$
m''_i : = \left\{
\begin{array}{ll}
m_i - 2L \al & ~~ i \in S,
\\
m_i & ~~ \mbox{otherwise}.
\end{array}
\right.
$$

On the other hand, the $\gamma$-twisting of 
$\cF^{(\sigma)}_I(\tau)$ is expressed as 
\begin{align}
\gamma_{(a,b)}\cdot \cF^{(\sigma)}_{\left\{(\ell_i, m_i) \right\}} (\tau)
= \left\{
\begin{array}{ll}
(-1)^{b\sum_{i\in S_1}\ell_i} \cF^{(\sigma)}_{\left\{(\ell_i, m_i) \right\}}(\tau), & ~~ (a =0)
\\
(-1)^{b\sum_{i\in S_1}(\ell_1+1)} \cF^{(\sigma)}_{\left\{(\ell'_i, m_i) \right\}}(\tau), & ~~ (a =1)
\end{array}
\right.
\label{gamma cF S}
\end{align}
where we set
$$
\ell'_i : = \left\{
\begin{array}{ll}
k_i- \ell_i & ~~ i \in S_1,
\\
\ell_i & ~~ \mbox{otherwise}.
\end{array}
\right.
$$

~


We next describe explicitly the Jacobi's theta functions twisted by the $\delta$-actions
given in \eqn{deltaL th}, that is,  
\begin{equation}
\th_{i, (\al,\beta)}(\tau) := 
q^{\frac{\al^2}{8}} e^{2\pi i \frac{\al \beta}{8}} \th_i\left(\tau, \frac{\al\tau+\beta}{2} \right),
\hspace{1cm} (i=3,4,2,1).
\label{th al beta}
\end{equation}

They are explicitly written down as follows;
\begin{description}
\item[(i) $\al,\beta \in 2\bz$ : ]
\begin{align}
& \th_{3, (\al,\beta)} (\tau) = (-1)^{\frac{\al \beta}{4}} \th_3(\tau), 
\nn
& \th_{4, (\al,\beta)} (\tau) = (-1)^{\frac{\al \beta}{4}+ \frac{\al}{2}} \th_4(\tau), 
\nn
& \th_{2, (\al,\beta)} (\tau) = (-1)^{\frac{\al \beta}{4}+ \frac{\beta}{2}} \th_2(\tau),
\nn
& \th_{1, (\al,\beta)} (\tau) = (-1)^{\frac{\al \beta}{4}+ \frac{\al+\beta}{2}} \th_1(\tau) \equiv 0, 
\end{align}

\item[(ii) $\al \in 2\bz, ~ \beta \in 2\bz+1 $ : ]
\begin{align}
& \th_{3, (\al,\beta)} (\tau) = e^{i\pi \left(\frac{\al \beta}{4} + \frac{\al}{2} \right)} 
\th_4(\tau), 
\nn
& \th_{4, (\al,\beta)} (\tau) =  e^{i\pi \frac{\al \beta}{4}}  \th_3(\tau), 
\nn
& \th_{2, (\al,\beta)} (\tau) =  e^{i\pi \left(\frac{\al \beta}{4} + \frac{\al+\beta+1}{2}\right)} 
\th_1(\tau) \equiv 0,
\nn
& \th_{1, (\al,\beta)} (\tau) =  e^{i\pi \left(\frac{\al \beta}{4} + \frac{\beta-1}{2} \right)} 
\th_2(\tau) , 
\end{align}

\item[(iii) $\al \in 2\bz+1, ~ \beta \in 2\bz $ : ]
\begin{align}
& \th_{3, (\al,\beta)} (\tau) = e^{i\pi \frac{\al \beta}{4}} \th_2(\tau), 
\nn
& \th_{4, (\al,\beta)} (\tau) =  e^{i\pi \left(\frac{\al \beta}{4} + \frac{\al}{2} \right)}  
\th_1(\tau)\equiv 0, 
\nn
& \th_{2, (\al,\beta)} (\tau) =  e^{i\pi \left(\frac{\al \beta}{4}+\frac{\beta}{2} \right)} 
\th_3(\tau) ,
\nn
& \th_{1, (\al,\beta)} (\tau) =  e^{i\pi \left(\frac{\al \beta}{4} + \frac{\al +\beta}{2}\right)} 
\th_4(\tau) , 
\end{align}

\item[(iv) $\al, \beta \in 2\bz+1$ : ]
\begin{align}
& \th_{3, (\al,\beta)} (\tau) = e^{i\pi \left(\frac{\al \beta}{4}+ \frac{\al}{2}\right)} 
\th_1(\tau) \equiv 0, 
\nn
& \th_{4, (\al,\beta)} (\tau) =  e^{i\pi \frac{\al \beta}{4}}  \th_2(\tau), 
\nn
& \th_{2, (\al, \beta)} (\tau) =  e^{i\pi \left(\frac{\al \beta}{4}+\frac{\al+\beta+1}{2} \right)} 
\th_4(\tau) ,
\nn
& \th_{1, (\al,\beta)} (\tau) =  e^{i\pi \left( \frac{\al \beta}{4} + \frac{\beta-1}{2} \right)} 
\th_3(\tau) .
\end{align}

\end{description}

~


\end{document}